# Jeans' criterion and nonextensive velocity distribution function in kinetic theory

Du Jiulin

*Department of Physics, School of Science, Tianjin University, Tianjin 300072, China*

*E-mail*: jiulindu@yahoo.com.cn   *or*   dujiulin@snnu.edu.cn

**Abstract**

The effect of nonextensivity of self-gravitating systems on the Jeans' criterion for gravitational instability is studied in the framework of Tsallis statistics. The nonextensivity is introduced in the Jeans problem by a generalized $q$-nonextensive velocity distribution function through the equation of state of ideal gas in nonextensive kinetic theory. A new Jeans' criterion is deduced with a factor $\sqrt{2/(5-3q)}$ that, however, differs from that one in Ref.[21] and new results of gravitational instability are analyzed for the nonextensive parameter $q$. An understanding of physical meaning of $q$ and a possible seismic observation to find astronomical evidence for a value of $q$ different from unity are also discussed.

The statistical mechanics of self-gravitating systems with long-range interaction exhibits very peculiar features such as negative specific heats [1-4] and the so-called gravothermal instability [5,6], which are greatly different from the usual thermodynamic systems and are therefore quite difficult to understand with the conventional theory. In recent years, it has been considered that systems with the gravitational long-range interaction may be nonextensive and, therefore, the conventional Boltzmann-Gibbs (B-G) statistical mechanics may be not appropriate to the description of the features of the systems. The nonextensive generalization of B-G statistical mechanics proposed by Tsallis has focused significant attention, which has been known as "Tsallis statistics"[7]. C. Tsallis made the nonextensive generalization of B-G entropy by introducing the following form of entropy [8]:

$$S_q = \frac{k}{1-q}(\sum_i p_i^q - 1) \tag{1}$$

where $k$ is the Boltzmann's constant, $p_i$ the probability that the system under consideration is in its $i$th configuration such that $\sum_i p_i = 1$, and $q$ a positive nonextensive parameter whose deviation from unity describes the degree of nonextensivity of the system. The celebrated B-G entropy $S_B$ is so obtained as the limit $q \to 1$:

$$S_q = \lim_{q \to 1} S_q = -k \sum_i p_i \ln p_i \tag{2}$$

The fundamental difference between the Tsallis entropy $S_q$ and the B-G entropy $S_B$ lies in that if the system is composed of two independent subsystems, $a$ and $b$, then the total Tsallis' entropy of the system satisfies $S_q(a+b) = S_q(a) + S_q(b) + (1-q)S_q(a)S_q(b)/k$. This new theory provides a convenient frame for the thermo-statistical analyses of many astrophysical systems and processes, such as stellar polytropes [9], the velocity distribution of galaxy clusters [10], thermodynamics of self-gravitating systems [11-13], the gravothermal catastrophe [5,6,14] and the long-term evolution [15] of self-gravitating systems in nonequilibrium, and the solar neutrino problem [16,17] etc. These literatures make it clear that Tsallis statistics may be the appropriate theory for description of astrophysical systems with the long-range interaction of gravitation.

In this letter, we analyze the effect of nonextensivity on the Jeans' criterion of gravitational instability by using a $q$-nonextensive velocity distribution function of kinetic theory that was proposed in the framework of Tsallis' statistics.

The dynamical stability of a self-gravitating system usually can be described by the Jeans' criterion of gravitational instability. The equations of the Jeans' problem are the equation of continuity, the Euler's equation, the Poisson's equation and the equation of state of an ideal gas (the isothermal gas)[18]:

$$\frac{\partial \rho}{\partial t} + \nabla(\rho \mathbf{v}) = 0 \tag{3}$$

$$\frac{\partial \mathbf{v}}{\partial t} + (\mathbf{v} \cdot \nabla)\mathbf{v} = -\frac{1}{\rho}\nabla P - \nabla \varphi \tag{4}$$

$$\nabla^2 \varphi = 4\pi G \rho \tag{5}$$

$$P = \frac{kT}{\mu m_H}\rho, \quad (T \text{ is constant}) \tag{6}$$

where, as usual, $\rho$ is mass density, $\mathbf{v}$ velocity of the fluid, $P$ pressure, $T$ temperature, $G$ gravitational constant, $\mu$ the mean molecular weight, $m_H$ the atomic mass of hydrogen, and $\varphi$ the gravitational potential. On the basis of these equations, the conclusion of Jeans' criterion is that density fluctuations with the wavelength $\lambda$ more than the critical value $\lambda_J = v_s \sqrt{\pi/G\rho_0}$ will grow so that the system becomes gravitationally unstable [18]. This means that a cloud of gas with the length scale more than $\lambda_J$ is gravitational instability and has to contract constantly. In this formula, $\lambda_J$ is the so-called Jeans' length, $v_s = \sqrt{kT/\mu m_H}$ is the speed of sound and $\rho_0$ is the undisturbed mass density.

The nonextensivity in the Jeans problem is introduced through the equation of state of an ideal gas. In the nonextensive kinetic theory, the $q$-nonextensive velocity distribution function for free particles is given through a nonextensive generalization of the Maxwell ansatz [19] by

$$f(v) = B_q \left[1 - (1-q)\frac{mv^2}{2kT}\right]^{\frac{1}{1-q}} \tag{7}$$

where $B_q$ is a normalization constant, $m$ and $v$ is mass and velocity of the particle, respectively, $k$ is the Boltzmann's constant. This theory has been used to analyze the negative heat capacity [20] and the instability [21] of a self-gravitating system, plasma oscillations [22] and the $H$ theorem for nonextensive transport equation [23].

We consider a cloud of ideal gas within the non-relativistic gravitational context. If $n$ denotes the particle number density, in the view of kinetic theory, pressure of the gas is defined by $P = \frac{1}{3}nm<v^2>$ with $<v^2>$ the mean square velocity of the particle. When we consider the nonextensive effect on the gas, the nonextensivity is introduced through a new expectation value of square velocity $<v^2>_q$ in Tsallis statistics that is defined [24] by

$$<v^2>_q = \frac{\int v^2 [f(v)]^q d^3v}{\int [f(v)]^q d^3v} \tag{8}$$

For the velocity distribution function (7), there is a thermal cutoff on the maximum value allowed for the velocity of the particle for $q<1$ [20-22]:

$$v_{max} = \sqrt{\frac{2kT}{m(1-q)}} \tag{9}$$

whereas for $q>1$ without the thermal cutoff, $v_{max} \to \infty$. In this way, the $q$ expectation value (8) for the square velocity of the particle is calculated [20] as

$$<v^2>_q = \frac{6}{5-3q}\frac{kT}{m}, \quad (0<q<\frac{5}{3}). \tag{10}$$

The result of this integral can also be obtained by the method of generalized equipartition in Ref [25]. It is clear that, as expected, the standard mean square velocity $<v^2>=3kT/m$ is correctly recovered from the above equation if we take $q \to 1$.

From Eq.(10) we directly obtain the equation of state of an ideal gas in the nonextensive kinetic theory

$$P = \frac{1}{3}nm<v^2>_q = \frac{2}{5-3q}nkT \tag{11}$$

By taking into account the particle mass $m = \mu\, m_H$, the particle number density $n = \rho / \mu\, m_H$, we can write Eq.(11) in the form of Eq.(6) as

$$P = \frac{2}{5-3q} \frac{k}{\mu\, m_H} \rho\, T \tag{12}$$

The standard equation of state of an ideal gas is recovered perfectly from the above two equations at the limit $q \to 1$.

We now analyze the Jeans problem determined by Eqs.(3), (4), (5) and (12) when the nonextensive effect is taken into consideration. As in the conventional way of stability analyses, we use $\rho_1$, $\mathbf{v}_1$, $\varphi_1$ and $P_1$ to be small perturbations in the system about its equilibrium state $\rho_0$, $\mathbf{v}_0$, $\varphi_0$ and $P_0$, respectively. Assume the conditions under which the system is static and homogeneous when it is at its equilibrium to be that $\mathbf{v}_0 = 0$, $\rho_0$ and $P_0$ are constant, respectively, then, let $\rho = \rho_0 + \rho_1$, $\mathbf{v} = \mathbf{v}_0 + \mathbf{v}_1$, $\varphi = \varphi_0 + \varphi_1$ and $P = P_0 + P_1$, we obtain the linearized equations as follows

$$\frac{\partial \rho_1}{\partial t} + \rho_0 \nabla \cdot \mathbf{v}_1 = 0 \tag{13}$$

$$\frac{\partial \mathbf{v}_1}{\partial t} = -\nabla \varphi_1 - \frac{1}{\rho_0} \nabla P_1 \tag{14}$$

$$P_1 = \frac{2}{5-3q} \frac{kT}{\mu\, m_H} \rho_1 \tag{15}$$

$$\nabla^2 \varphi_1 = 4\pi G \rho_1 \tag{16}$$

The amalgamation of these four equations yields the wave equation

$$\frac{\partial^2 \rho_1}{\partial t^2} = 4\pi G \rho_0 \rho_1 + \frac{2}{5-3q} \frac{kT}{\mu\, m_H} \nabla^2 \rho_1 \tag{17}$$

Eq.(17) has solutions in the form of $\rho_1 \sim \exp\left[\omega\, t + i(\frac{2\pi x}{\lambda})\right]$ with the relation between the frequency $\omega$ and the wavelength $\lambda$ as

$$\omega^2 = 4\pi G\rho_0 - \left(\frac{2\pi}{\lambda}\right)^2 \frac{2}{5-3q}\frac{kT}{\mu m_H} \tag{18}$$

The solution of this equation leads to a critical wavelength $\lambda_C$ for the instability

$$\lambda_C = \sqrt{\frac{2}{5-3q}\frac{\pi kT}{\mu m_H G\rho_0}} = \lambda_J \sqrt{\frac{2}{5-3q}} \tag{19}$$

where $\lambda_J$ is known as Jeans' length. Eq.(19) gives the conclusion that if the wavelength $\lambda$ of density fluctuation is more than the critical one $\lambda_C$, the density will grow with time in the exponent form and the system will become gravitationally unstable. It is clear that the criterion of gravitational instability depends on the nonextensive parameter $q$. Thus, the Jeans' criterion is modified due to the nonextensive effect by

$$\lambda > \lambda_J \sqrt{\frac{2}{5-3q}} \tag{20}$$

with a factor $\sqrt{2/(5-3q)}$ and $q$ in the value range of $0 < q < 5/3$.

In conclusion, we have studied the effect of nonextensivity on the Jeans' criterion. The nonextensivity is introduced in the Jeans' problem through the equation of state of an ideal gas by the $q$ velocity distribution function (7) in the nonextensive kinetic theory. We obtain a new criterion for gravitational instability from (19) and (20). This nonextensive modification of the Jeans' criterion leads to a new critical length $\lambda_C$ that depends on the nonextensive parameter $q$ as follows

(a) If $q = 1$, the Jeans' criterion is recovered perfectly, $\lambda > \lambda_C = \lambda_J$.

(b) If $0 < q < 1$, the Jeans' criterion is modified as $\lambda > \lambda_C < \lambda_J$.

(c) If $1 < q < 5/3$, the Jeans' criterion is modified as $\lambda > \lambda_C > \lambda_J$.

(d) If $q \to 5/3-$, $\lambda_C \to \infty$, the self-gravitating system is always stable.

From above results we conclude that, for the gas with $0 < q < 1$, the critical length is decreased by the factor $\sqrt{2/(5-3q)}$ and the system may present gravitational instability even if wavelengths of the density fluctuation is less than the Jeans' length $\lambda_J$. This means that a cloud of gas with $0< q <1$, which was believed to be stable gravitationally on the basis of the standard B-G statistics, may be now unstable according to Tsallis' statistics. However, for the gas with $1< q <5/3$, the critical length is increased by the factor $\sqrt{2/(5-3q)}$ and the system may remain stable even if wavelengths of the density fluctuation is more than the Jeans' length $\lambda_J$. It is clear that a cloud of gas with $1< q <5/3$, which was believed to be unstable gravitationally on the basis of the standard B-G statistics, now may remain stable according to Tsallis' statistics.

A recent work [21] on the Jeans' gravitational instability and nonextensive kinetics has attracted our attentions. These authors used the velocity distribution function (7) to analyze a nonextensive modification of the Jeans' criterion from a kinetic approach based on the Vlasov equation, which leads to a critical wavelength for gravitational instability $\lambda_C = \lambda_J \sqrt{2/(3q-1)}$, ($q>1/3$), with a factor $\sqrt{2/(3q-1)}$ that differs from our result (19). However, if the nonextensive velocity distribution function uses Eq.(7) but the mean square velocity of the particle uses its standard definition instead of Eq.(8), we find [22]

$$<v^2> = \int v^2 f(v) d^3v = \frac{6}{3q-1}\frac{kT}{m}, \quad (q>\frac{1}{3}). \tag{21}$$

and then, the equation of state of an ideal gas becomes

$$P = \frac{1}{3}nm<v^2> = \frac{2}{3q-1}nkT \tag{22}$$

If we solve the Jeans problem that is determined by Eqs.(3), (4),(5) and (22), we can obtain the critical wavelength of Ref.[21]. However, the choice (21) for the standard definition of expectation value is inadequate for the Tsallis statistics [24] and seems to have been discarded.

How to understand the physical meaning of the nonextensive parameter $q$ plays a very important role in the applications of Tsallis statistics to the fields of astrophysics. But, in the light of present understanding, it is still an open problem. With the standard B-G statistics, the structure and stability of self-gravitating systems at statistical equilibrium are analyzed usually in terms of the maximization of a thermodynamic potential (the so-called mean field description). This thermodynamic approach leads to isothermal configurations (for example, self-gravitational, isothermal gas spheres) satisfying the equation of state of Eq.(6) that have been studied for long time in the stellar structure and the galactic structure [11,12]. It is well known that isothermal configurations only correspond to meta-stable states (locally mixing), not true equilibrium states. When the nonextensive effect is considered in the framework of Tsallis statistics, the equation of state of an idea gas can be written as $P_q = nkT_q$ with the physical temperature $T_q$ a variable [26] that depends on the nonextensive parameter $q$. Compare this equation with Eq.(11), we find $T_q = 2T/(5-3q)$. In this way, we agree with the explanation for $q$ of self-gravitating systems that $q=1$ represents an isothermal processes of the gas, corresponding to the state of complete mixing, but $q \ne 1$ is nonisothermal, corresponding to the state of incomplete mixing, it measures the degree of mixing. However, we are more inclining to that $q$ values of a self-gravitating system might be *space inhomogeneous*, i.e., the degree of mixing might depend on the space coordinate.

Recent investigations of nonextensive effects on self-gravitating systems lead to the power-law distribution referred to as the stellar polytrope [5, 6, 9,14], which relates $q$ to the polytrope index n as n = 3/2 + 1/($q$-1) depending on the standard statistical averages [5, 9] or n =1/2+ 1/(1-$q$) depending on the normalized $q$ averages [14]. It is noticeable that the limit n $\to \infty$ (or q $\to$ 1) is corresponding to the isothermal distribution derived from the B-G statistics. By performing a set of numerical simulation of long-term stellar dynamical evolution [15], it has been found that this quasi-equilibrium sequence with different $q$ values arising from the Tsallis entropy plays an important role in characterizing the transient state away from the B-G equilibrium state in the nonequilibrium dynamical evolution of stellar self-gravitating systems.

It has been reported [10] that the observed velocity distribution of galaxy clusters significantly deviates from the Maxwellian distribution and can be fitted well by the Tsallis nonextensive distribution with $q = 0.23^{+0.07}_{-0.05}$. In addition, Tsallis statistics could give the speed of sound, $v_{s_q} = \sqrt{kT_q/\mu m_H}$, different from that one in B-G statistics ($q$ =1, $T_q$= $T$), with the physical temperature $T_q$, for example, $T_q = 2T/(5-3q)$ in the gas model of this Letter. The different speeds of sound predicted by the tow statistics cause a possibility to find the experimental evidence for the nonextensive effect. As a means of probing the interior structure and dynamics of a star with increasing precision, the helioseismology has provided the information about the square speed of sound through stellar interiors [27]. With this seismic observation, in principle, we might find astronomical or experimental evidence for a value of *q* different from unity.

**Acknowledgments**

The author would like to thank Professor R. Silva for his helpful discussions of the *q*-nonextensive velocity distribution and Professor J. A. S. Lima for his recent reprint. This work is supported by the project of "985" Program of TJU of China.